\documentclass[prc,twocolumn,nofootinbib,showpacs,floatfix,letterpaper]{revtex4}
\usepackage{latexsym}
\usepackage{graphicx,slashbox}
\usepackage{amsmath,amssymb,multirow}
\usepackage[body={18cm,23cm},top=2.5cm]{geometry}
\renewcommand{\arraystretch}{1.2}
\newcommand{\eq}[1]{\begin{equation} #1 \end{equation}}
\newcommand{\ac}[1]{a^{\dagger}_{#1}}

\newcommand{\elmx}[3]{\langle #1|#2|#3 \rangle}
\newcommand{\ket}[1]{|#1\rangle}

\newcommand{\HHF}{\hat{H}_{\mathrm{HF}}}
\newcommand{\Hqp}{\hat{H}_{\mathrm{IQP}}}
\newcommand{\VHF}{\hat{V}_{\mathrm{HF}}}
\newcommand{\Vres}{\hat{V}_{\mathrm{res}}}

\begin{document}


\title{Ground-state properties of even-even $N=Z$ nuclei within the 
Hartree--Fock--BCS and Higher Tamm--Dancoff approaches}

\author{L. Bonneau\textsuperscript{1}}
\author{P. Quentin\textsuperscript{2}}
\author{K. Sieja\textsuperscript{2,3}}

\affiliation{\textsuperscript{1}\mbox{Theoretical Division, 
Los Alamos National Laboratory, Los Alamos, New Mexico 87545, USA}\\
\textsuperscript{2}Centre d'Etudes Nucl\'eaires de 
Bordeaux-Gradignan, CNRS-IN2P3 and Universit\'e Bordeaux-I, BP 120, 
33175 Gradignan, France\\
\textsuperscript{3}Institute of Physics, University of 
M. Curie-Sk{\l}odowska, ul. Radziszewskiego 10, 20-031 Lublin, Poland}

\date{\today}

\begin{abstract}

We calculate the ground-state properties of well deformed, 
even-even $N=Z$ nuclei in the region between $^{56}$Ni and $^{100}$Sn 
within two different approaches, focusing on the binding energy and 
deformation and pairing properties. First, we employ 
the Hartree--Fock--BCS (HFBCS) approximation with the Skyrme effective 
nucleon-nucleon interaction and discuss how the results depend on the 
parameterization of the interaction and on the pairing force parameters 
adjusted in various schemes to reproduce the experimental 
odd-even mass differences. Then, within the Higher 
Tamm--Dancoff Approximation (HTDA), which explicitly conserves 
the particle number, we calculate the same properties starting from 
the HFBCS solutions. The HTDA treatment of the ground-state correlations 
is converged within a $n$-particle--$n$-hole expansion using up to $n$=4 
particle-hole excitations of the pair type (in the sense of Cooper pairs). 
We compare the ground-state properties calculated in these two descriptions 
of pairing correlations and deduce the importance of the particle-number 
conservation in weak pairing regimes. Finally, we extend the HTDA 
calculations so as to include the proton-neutron residual interaction 
and investigate the role of proton-neutron pairing on the above 
ground-state properties. 
 
\end{abstract}


\maketitle

%
%

\section{Introduction}

Nuclei with an equal number of neutrons and protons are of a special 
interest in many respects. The similarity of the neutron and proton 
single-particle (sp) spin-space wave functions, in the vicinity of 
the chemical potentials, allows for rather interesting physical studies 
associated with isospin mixing, $\beta$ decays and proton-neutron  
correlations. Such peculiar correlations are encountered in pairing 
properties either in the $T = 0$ or $T = 1$ channels. Another 
intriguing property is the so-called Wigner term in a liquid-drop 
approaches, corresponding to a very sharp pattern of extra stability 
close to $N = Z$. These nuclei are also of importance for the 
rapid-proton astrophysical process. Up to $A \sim 60$ for even values 
of $N=Z$, these nuclei are very close indeed to the proton drip line. 
However, contrary to the situation near the neutron drip line and due 
to the Coulomb barrier, their description in a 
mean-field-plus-correlations approach is not marred by the 
necessity of dealing 
accurately with the treatment of the continuum. If one is interested 
in specific properties of such nuclides and their neighbors (such as 
the Wigner term), it is desirable to use a description of pairing 
correlations which is not blurred by fluctuations of the 
neutron and proton numbers as is the case when using Bogoliubov 
quasiparticle vacua as ansatz for the ground-state (GS) wave functions.

In this paper we are concerned with the mean-field and beyond-mean-field 
descriptions of well-deformed even-even $N=Z$ nuclei lying between the 
doubly magic nuclei $^{56}$Ni and $^{100}$Sn. To do so we use various 
Skyrme effective interactions (SIII~\cite{Beiner75} and SLy4~\cite{Chabanat98}) 
supplemented by two different treatments of the pairing 
correlations: \`a la Bogoliubov-Valatin (BCS wave function) and within 
the particle-number conserving approach dubbed as the Higher Tamm--Dancoff
Approximation (HTDA)~\cite{Pillet02}, in both cases for like-particle 
pairing correlations. Then we add in the latter treatment 
proton-neutron pairing correlations in $T = 0$ or $T = 1$ isospin 
channels.

In view of its widespread use for several decades, the BCS approach 
for like-particle pairing properties is not reviewed here. 
Only some practical details (among which the choice of the relevant 
average matrix elements is of paramount importance) are discussed. 
In contrast the HTDA approach is less widely known and is briefly 
described here. 

The HTDA approach may be presented as a treatment of correlations 
in a highly truncated shell model whose practicability and 
efficiency rely on the fast convergence of the particle-hole expansion. 
This has been shown to be realized upon choosing for the particle-hole
quasiparticle vacuum a relevant Hartree--Fock (HF) solution associated
self-consistently with the one-body reduced density matrix of the 
correlated wave function. The HTDA approach was applied for the first time 
to describe the ground and isomeric states in $^{178}$Hf~\cite{Pillet02} 
and then odd nuclei and most general isomeric states \protect\cite{Ha04}. 
A Routhian-HTDA scheme was then proposed to describe the superdeformed 
yrast bands in the $A\sim190$ region~\cite{Quentin04}. A preliminary 
HTDA study of the GS pairing correlations in $^{64}$Ge, 
including isovector and isoscalar residual interactions, was presented 
in Ref.~\cite{Sieja07}. Here the HTDA approach is applied for the first 
time in systematic calculations of the properties of medium-mass, 
proton-rich nuclei.

The outline of the present work is the following. In Sect.~\ref{form}
we describe the theoretical background (mostly the HTDA formalism) 
and explain at length in Sec.~\ref{calculation_procedure} 
how the calculations are carried out in practice, especially 
regarding the pairing strength fitting strategy in both BCS and 
HTDA approaches and the optimization of the harmonic-oscillator 
basis parameters in the \mbox{HFBCS} calculations. The presentation 
and discussion of the obtained results is organized in three steps. 
First, in Sec.~\ref{res_HFBCS}, we focus on the GS 
properties obtained within the HFBCS approach, and in Sect.~\ref{res_HTDA} 
on the results obtained within the HTDA formalism without proton-neutron 
coupling in the particle-particle channel. Then we carry out in 
Sect.~\ref{sec-pair} a comparative study of pairing properties depending 
on the pairing treatment and the fitting scheme. Finally we extend 
in Sec.~\ref{pn_pairing} the HTDA calculations so as to include the 
full isovector and isoscalar pairing interaction. The main conclusions 
of our study are drawn in Sec.~\ref{conclusion}.

%
%

\section{Theoretical background}

\label{form}

In the HFBCS approach we use the Skyrme effective interaction in the 
particle-hole channel and the seniority force in the particle-particle 
channel, and expand the single-particle wave functions in the cylindrical 
harmonic-oscillator basis, as detailed in Ref.~\cite{Flocard73}. 

We focus in the remainder of this section on the HTDA approach. Its 
purpose is to describe various nucleon correlations (such as pairing 
and RPA) on the same footing. Among the many particle-hole excitations 
on a Slater determinant vacuum (noted here $|\Phi_0\rangle$\footnote{For 
the sake of clarity in the notation, we reserve the letter $\Phi$ for 
a Slater determinant and the letter $\Psi$ for a correlated state.}), 
pair excitations around the Fermi surface play an essential role. 
By construction the HTDA approach is an extension of the Tamm--Dancoff 
approximation to higher order of particle-hole excitations, so it may 
be regarded as a truncated shell model. The rapidity of the 
convergence of the particle-hole expansion, and thus the 
tractability of the approach, depends on the realistic 
character of $|\Phi_0\rangle$. A fast 
convergence is expected to be reached when the quasi-vacuum is defined 
self-consistently in such a way that some many-body effects of 
the correlations are already taken into account at the mean-field level.

Let us consider the effective Hamiltonian
\eq{
\label{Htot}
\hat{H}=\hat{K}+\hat{V}\,,
}
where $\hat{K}$ denotes the kinetic energy operator and $\hat{V}$ an 
effective interaction. For the wave function $\ket{\Phi_0}$ we choose 
the HF solution, i.e., the eigenstate of the Hamiltonian $\HHF$ 
defined by
\eq{
\HHF=\hat{K}+\VHF
}
where the potential $\VHF$ denotes the one-body reduction 
of $\hat V$ for $\ket{\Phi_0}$ and is self-consistently 
obtained from the many-body reduced density matrix $\hat{\rho}$ of 
the correlated solution $\ket{\Psi}$ for the desired number of 
particles. We have thus
\eq{
\hat H_{\rm HF}\ket{\Phi_0}=E_0 \ket{\Phi_0}\,,
}
where $E_0$ is the associated eigenenergy and
\eq{
\hat H_{\rm HF}=\hat{K}+\VHF\,.
}
This approach also allows to include various one-body 
constraints (on the nuclear deformation for example) in a simple 
way since the constraint operator can be absorbed in the definition 
of $\HHF$. In addition we assume here that the GS solutions 
possess the time-reversal, axial and parity symmetries.

The quasi-vacuum $\ket{\Phi_0}$ may now serve to construct an orthonormal 
many-body basis in which we diagonalize the Hamiltonian $\hat{H}$. 
In principle to build this basis we should include, besides 
$\ket{\Phi_0}=\ket{\Phi_0^\tau}\otimes\ket{\Phi_0^{\tau'}}$, 
the particle-hole excitations of all orders (from 1 to the 
particle number) created on $\ket{\Phi_0}$, noted generically 
$\ket{\Phi_n}$ for $n$-particle--$n$-hole ($n$p-$n$h) excitations. 
The total GS wave function $\ket{\Psi}=\ket{\Psi^\tau}\otimes
\ket{\Psi^{\tau'}}$ can therefore be decomposed in the following way
\eq{
\begin{aligned}
\label{mbstpn}
& \ket{\Psi} = \chi_{00}\ket{\Phi_0^\tau}\otimes\ket{\Phi_0^{\tau'}} 
+\sum_{(1{\rm p}-1{\rm h})_\tau}\chi_{10}
\ket{\Phi_1^\tau}\otimes\ket{\Phi_0^{\tau'}} \\
& +\sum_{(1{\rm p}-1{\rm h})_{\tau'}}\chi_{01}
\ket{\Phi_0^\tau}\otimes\ket{\Phi_1^{\tau'}} 
+\sum_{\substack{(1{\rm p}-1{\rm h})_{\tau}\\(1{\rm p}-1{\rm h})_{\tau'}}}
\chi_{11}\ket{\Phi_1^\tau}\otimes\ket{\Phi_1^{\tau'}} \\
& +\sum_{(2{\rm p}-2{\rm h})_\tau}\chi_{20}
\ket{\Phi_2^\tau}\otimes\ket{\Phi_0^{\tau'}}
+\sum_{(2{\rm p}-2{\rm h})_{\tau'}}\chi_{02}
\ket{\Phi_0^\tau}\otimes\ket{\Phi_2^{\tau'}}\\
& + \cdots
\end{aligned}
}
where $\tau$ and $\tau'$ denote two different charge states. However 
practical calculations require to truncate this expansion. Based on 
the former studies in the HTDA framework~\cite{Pillet02,Ha04,Sieja07}, 
we may assume that the components of the pair-excitation type dominate 
in the GS solution. The set of products of Slater determinants 
$\ket{\Phi_i^\tau}\otimes\ket{\Phi_j^{\tau'}}$ is an orthonormal basis 
of the physical space accessible to a nucleus having $N$ neutrons and 
$Z$ protons. Assuming time-reversal symmetry, the coefficients 
$\chi_{i}$ in Eq.~(\ref{mbstpn}) are real and, if we take $\ket{\Psi}$ 
normalized to unity, obey the relation
\begin{equation}
\sum_{i} \chi_{i}^2=1\,.
\end{equation}
It can be easily shown that the expression~(\ref{mbstpn}) of 
$\ket{\Psi}$ ensures that $\ket{\Psi}$ is an eigenstate of the 
particle-number operator: $\hat N\ket{\Psi}=A\ket{\Psi}$, with $A=N+Z$. 
Finally, to obtain the correlated ground state we diagonalize the 
Hamiltonian defined in Eq.~(\ref{Htot}) in the retained many-body 
basis.

Now, let us rewrite the Hamiltonian $\hat{H}$ as
\eq{
\hat{H}=\elmx{\Phi_0}{\hat{H}}{\Phi_0}+\Hqp+\Vres\:,
}
where the independent quasiparticle Hamiltonian $\Hqp$ and 
the residual interaction $\Vres$ are defined by
\begin{gather}
\Hqp=\HHF-\elmx{\Phi_0}{\HHF}{\Phi_0}\:,\\
\Vres=\hat V-\hat{\mathcal{V}}_{\rm HF}+\elmx{\Phi_0}{\hat{V}}{\Phi_0}\:.
\end{gather}
It is interesting to note that these definitions give a vanishing 
expectation value of $\Hqp$ and $\Vres$ for the HF solution $\ket{\Phi_0}$. 
The independent quasiparticle Hamiltonian can also be expressed as
\eq{
\label{Hqp}
\Hqp=\sum_k \xi_k\eta^{\dagger}_k\eta_k
}
where $\xi_k$ is equal to the energy $\epsilon_k$ of the single-particle 
state $\ket{k}$ if $\ket{k}$ is a particle state with respect to 
$\ket{\Phi_0^\tau}$ or equal to $-\epsilon_k$ if $\ket{k}$ is a hole 
state. In Eq.~(\ref{Hqp}), $\eta^{\dagger}_k$ is the creation 
operator $\ac{k}$ associated with $\ket{k}$ if $\ket{k}$ is a 
particle state or the annihilation operator $a_k$ if $\ket{k}$ is a 
hole state. The matrix element of $\hat{H}$ between two Slater 
determinants $\ket{\Phi_i}$ and $\ket{\Phi_j}$ of the multi-particle 
multi-hole basis therefore takes the form 
\eq{
\elmx{\Phi_i}{\hat{H}}{\Phi_j}=\delta_{ij}
\bigl(\elmx{\Phi_0}{\HHF}{\Phi_0}+\sum_{\tau}E_{\rm ph}^{i(\tau)}\bigr)
+\elmx{\Phi_i}{\Vres}{\Phi_j},
\label{hijkl}
}
where $E_{\rm ph}^{i(\tau)}$ is the particle-hole excitation energy 
associated with $\ket{\Phi_i^{\tau}}$ and calculated with respect to 
the vacuum $\ket{\Phi_0^\tau}$ as
\eq{
\label{Eph}
E_{\rm ph}^{i(\tau)}=\sum_{k\in \ket{\Phi_i^{\tau}}} \xi_k\:.
}
where the sum runs over all the single-particle states
$\ket{k}$ contained in $\ket{\Phi_i^{\tau}}$. Since the residual 
interaction contains only one- and two-body operators, the matrix element 
$\elmx{\Phi_i}{\Vres}{\Phi_j}$ vanishes when $\ket{\Phi_i}$ and 
$\ket{\Phi_j}$ differ by particle-hole excitations of order three 
or higher. It can be calculated in terms of two-body matrix elements 
of $\hat{V}$ by application of the Wick theorem generalized to 
quasi-vacua as shown in Ref.~\cite{Ha04}.


%
%

\section{Calculation procedure}

\label{calculation_procedure}

Since we are in a region of nuclear deformation instability (shape   
coexistence) our final results can be sensitive to the choice 
of the effective interaction and to the pairing treatment. This is why 
two parameterizations of the phenomenological Skyrme interaction,   
namely the SIII and SLy4 ones are used. In the particle-particle 
channel a simple seniority ansatz is adopted in the BCS calculations. 
It is specified by the value $G$ of the constant pairing matrix elements   
between any single-particle states in the canonical basis retained 
in this part of the calculations. The value of $G$ is adjusted to 
reproduce the so-called empirical pairing gaps evaluated 
in the 3-point or a 5-point formula \cite{Madland88}. The
differences between these two experimentally deduced quantities 
are large in this region and so are the experimental errors for nuclear 
masses. In addition the meaning of the pairing gap derived from   
finite mass differences is no longer clear in the $N=Z$
case. Therefore, we find it necessary to make a comparison of 
the results obtained in the two cases which represent stronger
(5-point fit) and somewhat weaker (3-point fit) pairing regimes. 
We describe in the next subsection the fitting procedure for $G$ as well as 
the harmonic-oscillator (HO) basis parameters optimization carried out
in the HFBCS framework. The obtained basis parameters are then used in
the HTDA calculations assuming that they do not differ significantly
from the values that would result from an optimization in the HTDA 
approach. 

Once the solutions corresponding to the equilibrium 
deformations determined in the HFBCS calculations are found, we perform 
perturbative HTDA calculations (one diagonalization of the HTDA matrix 
is performed on top of the HFBCS calculations) without any constraints 
on the deformation. Since we are interested in the GS correlations of
even-even nuclei, the many-body basis includes here only 
pair excitations, i.e., excitations where nucleons occupying 
twofold Kramer-degenerate hole levels are scattered into 
Kramer-degenerate particle levels. As for the residual interaction 
in the HTDA approach, we choose a $\delta$ force whose 
strength is adjusted as explained in subsection~\ref{HTDA_fit}.  
In this way our formalism not only retains the realistic character of
self-consistent mean-field calculations, but also ensures the
particle number conservation.
  
\subsection{Pairing strength adjustment in the HFBCS approach}

We shall follow here the steps of Bonche and collaborators 
\cite{Bonche85} and apply the BCS approximation 
with the seniority pairing interaction. For the sake of defining our
notation and for completeness, we recall the relevant expressions
involved, omitting the isospin $\tau$ index. First, the seniority 
antisymmetrized matrix element is given by
\eq{ 
\widetilde{V}_{k l k' l'}=-G\,f_kf_{k'}
\delta_{l\overline{k}}
\delta_{l'\overline{k'}}
} 
with the smooth cut-off factor
\eq{
\label{smooth_cut_off}
f_k=
\frac{1}{1+\exp\bigl[(\epsilon_k-\epsilon_F-\Delta\epsilon)/\mu\bigr]}\:. 
} 
Here as well as in the HTDA case, the Fermi level $\epsilon_F$ is defined by
\eq{ 
\epsilon_F=\frac{1}{2}\,(\epsilon_n+\epsilon_{n+1}) 
} 
where $\epsilon_n$ and $\epsilon_{n+1}$ denote the energies of the
last occupied and the first empty single-particle state, respectively
(in a pure HF picture), and $\Delta\epsilon$ and $\mu$ denote the
cut-off energy and the diffuseness parameters. The pairing gap can be 
expressed as
\eq{ 
\Delta_k=f_k\,\Delta 
} 
where the state independent gap $\Delta$ is given by
\eq{ 
\label{Delta} 
\Delta=\frac{G}{2}\sum_{k>0}\frac{f_k^2\Delta}{E^{(k)}_{\rm qp}}
}
with the quasiparticle energy $E^{(k)}_{\rm qp}$ defined by
\eq{
\label{Eqp} 
E^{(k)}_{\rm qp}=\sqrt{\widetilde{\epsilon}_k^2+f_k^2\Delta^2}\:.
} 
In Eq.~(\ref{Delta}), the sum runs over all the pairs of Kramer 
degenerate single-particle states, but in fact the cut-off factor 
$f_k$ suppresses the contributions of single-particle states lying 
at least about $\Delta\epsilon+\mu$ above the Fermi level. 
From the average particle number conservation
\eq{ 
2\sum_{k>0}v_k^2=N \:,
} 
we can deduce the expression of the chemical potential~$\lambda$
\eq{ 
\label{lambda} 
\lambda=\frac{N-\sum\limits_{k>0}\left(1-\frac{\epsilon_k}{E^{(k)}_{\rm qp}} 
\right)}{\sum\limits_{k>0}\frac{1}{E^{(k)}_{\rm qp}}}\:. 
} 
Finally the pairing energy takes here the simple form
\eq{ 
E_{\rm pair}=-\frac{\Delta^2}{G} \:. 
} 

We now discuss the determination of the pairing strengths
$G_0^{(\tau)}$ related to the actual 
matrix elements $G^{(\tau)}$ through the following prescription
\eq{ 
G^{(\tau)}=\frac{G_0^{(\tau)}}{11+N_{\tau}}\:. 
} 
First we approximately take into account the Coulomb reduction of
proton pairing by assuming that, as the Hartree--Fock--Bogoliubov 
(HFB) calculations with the Gogny D1S interaction tend to 
indicate~\cite{Anguiano01}:
\eq{ 
\label{antipairing} 
G_0^{(p)}=0.9\,G_0^{(n)}\:. 
} 
Then, for a given parameterization of the Skyrme force and a given set of 
basis parameters ($N_0$, $b$ and $q$ in the notation of 
Ref.~\cite{Flocard73}), we determine the GS deformation 
$\beta_2$ of each nucleus using a reasonable initial $G$-value, 
assumed to be the same for all the nuclei under study. 
Then we deduce $G= G_0^{(n)}$ from a least-square fit to the
experimental minimal quasiparticle energies through the 3-point and 5-point 
formulae~\cite{Madland88,Dobaczewski01} and using the same
single-particle spectrum (the one for the charge state $\tau$
corresponding to the converged solution at $\beta_2$). We thus have to
minimize the following function
\eq{
\chi^2(G)=\frac{1}{2\,N_{\rm nucl}}\sum_{i=1}^{N_{\rm nucl}}\sum_{\tau=n,p} 
\biggl(\bigl[\mathcal{E}_\tau(G)\bigr]_i-
\bigl[\Delta_{\tau}^{(\rm exp)}\bigr]_i\biggr)^2\,,
}
where $\mathcal{E}_{\tau}$ denotes the lowest quasiparticle energy of
the nucleons of type $\tau$ and $i$ is an index running over the
$N_{\rm nucl}$ nuclei included in the fit. With the obtained
$G$-value, we determine the new GS deformation of 
each nucleus and minimize again $\chi^2(G)$ to find a new $G$-value. 
This procedure is repeated until the simultaneous convergence of $G$
and the GS deformations $\beta_2^{(i)}$ is reached. 
In practice, it is necessary to scan a wide range of deformations to
find the lowest local minimum of the deformation energy curve. The
latter is determined using the basis parameters 
$b=\sqrt{m\omega_0/\hbar}$, with 
$\omega_0=(\omega_{\bot}^2\omega_z)^{1/3}$, and 
$q=\omega_{\bot}/\omega_z$ deduced from an approximate expression for
a HO potential (see Ref.~\cite{Flocard73}). This approximate
optimization requires only the knowledge of $b_0$, 
the optimized value of $b(q)$ for a spherical solution. 
We carry out this study with $b_0=0.505$, which is
approximately the actual optimal value for the considered nuclei. 
By varying the basis parameters in the calculated ground states of all
nuclei, we have checked that the optimal $G$-value does not change
significantly. The obtained values of $G$ are reported in
Table~\ref{Exp_Eqp}. 
\begin{table}[h]
\caption{Empirical pairing gaps deduced from the 
odd-even mass differences (using the Atomic Mass Evaluation AME2003 
\protect\cite{Audi03}) in 5-point and 3-point formulae  
\protect\cite{Madland88,Dobaczewski01}, except for 
$\Delta_p^{(5)}(^{80}\rm Zr)$ for which we use the nuclear mass 
calculated by M\"oller and Nix~\protect\cite{Moller95}. The
standard estimate of the pairing gap $12/\sqrt{A}$ is given for 
comparison. In the last column the experimental binding energy per 
nucleon is also indicated. All values are in MeV.\label{Exp_Eqp}}
\begin{center}
\renewcommand{\arraystretch}{1.8}
\begin{tabular}{cccccccc}
\hline
\hline
& Nucleus & $\Delta_n^{(5)}$ & $\Delta_p^{(5)}$ & $\Delta_n^{(3)}$ &
$\Delta_p^{(3)}$ & $12/\sqrt{A}$ &$E/A$ \\
\hline
& $^{64}$Ge & 2.07 & 1.85 & 1.48 & 1.14& 1.50 &8.5294\\
& $^{68}$Se & 2.24 & 2.01 & 1.66 & 1.37& 1.45 &8.4773\\
& $^{72}$Kr & 1.72 & 1.73 & 1.24 & 1.10& 1.41 &8.4293\\
& $^{76}$Sr & 1.46 & 1.635& 0.88 & 1.07& 1.37 &8.3938\\
& $^{80}$Zr & 1.94 & 1.57 & 1.31 & 0.99& 1.34 &8.3741\\
\hline
\hline
\end{tabular}
\end{center}
\end{table}

\subsection{Pairing strength adjustment in the HTDA approach}

\label{HTDA_fit}

As mentioned earlier, in the HTDA calculations we use a $\delta$ force 
in the particle-particle channel. Both the coupling constant and 
the cut-off parameter are necessary to define fully the interaction. 
In our approach it is then necessary to fix the strength of this 
interaction and this is done by adjusting it so as to reproduce 
physical quantities for the considered nuclei, e.g., the phenomenological 
gaps. For that purpose, assuming that the appearance of the pairing 
gap is related to the breaking of the Cooper pair of lowest energy, 
we block in the HTDA calculations the single-particle level (neutron 
or proton) closest to the Fermi energy. Then, we adopt the difference 
between the expectation value of $\Vres$ in a normal (n) and blocked 
(b) calculations as a proper measure of the pairing correlations 
that can be compared to the experimental odd-even mass differences. 
Namely, we define 
\begin{equation}
\Delta=\bigl(E^{(\rm n)}-E_{\rm IQP}^{(\rm n)}\bigr)-
\bigl(E^{(\rm b)}-E_{\rm IQP}^{(\rm b)}\bigr)\,,
\end{equation}
where $E=\elmx{\Psi}{\hat{H}}{\Psi}$ and 
$E_{\rm IQP}=\elmx{\Psi}{\Hqp}{\Psi}$.

%
%

\section{Results of the Hartree--Fock--BCS calculations} 

\label{res_HFBCS} 

The five selected well-deformed $N=Z$ nuclei are $^{64}$Ge, $^{68}$Se, 
$^{72}$Kr, $^{76}$Sr and $^{80}$Zr. For one of the two considered 
effective interactions (SIII) we use several pairing windows 
with different values of $\Delta\epsilon$ (6, 8 and 10~MeV) and 
$\mu$ (0.2 and 0.5~MeV). In the fitting process, whereas the choice 
of $\Delta\epsilon$ is rather unimportant in the considered range, 
we find that it is not the case for $\mu$. For instance, with 
$\mu=0.5$~MeV and $\Delta\epsilon=6$ MeV, the iterative procedure 
to adjust of $G$ undergoes oscillations preventing to reach convergence, 
contrary to all the other pairing windows considered here. 
We thus retain for further calculations the values $\Delta\epsilon=6$MeV 
and $\mu=0.2$MeV stemming as the best choice from the fit to the 5-point 
experimental gaps. The optimal pairing strengths $G_{\rm opt}$ for the 
different Skyrme interactions and fitting schemes are displayed in 
Table~\ref{Gopt} together with the root-mean-square error on the 
experimental quasiparticle energies $\sigma_{\Delta}$
\eq{ 
\sigma_{\Delta}=\sqrt{\chi^2(G_{\rm opt})} \:.
}
\begin{table}[h] 
\caption{Optimal $G$-values and corresponding root-mean-square error 
on the experimental gaps $\sigma_{\Delta}$ (in MeV) obtained with the 
SIII and SLy4 Skyrme interactions using the 3-point and the 5-point 
formulae with $\Delta\epsilon=6$~MeV, $\mu=0.2$~MeV and $N_0=10$. 
\label{Gopt}} 
\begin{center} 
\begin{tabular}{cccccc} 
\hline\hline 
& Force & Formula & $G_{\rm opt}$ & $\sigma_{\Delta}$ &  \\ 
\hline 
& SIII & 3-point & 17.7 & 0.120 & \\ 
& SIII & 5-point & 20.6 & 0.216 & \\ 
& SLy4 & 3-point & 17.2 & 0.194 & \\ 
& SLy4 & 5-point & 19.9 & 0.260 & \\ 
\hline\hline 
\end{tabular} 
\end{center} 
\end{table} 

These adjustments have been performed with a basis size defined by 
$N_0=10$. The same adjustment procedure has also been carried out with 
a much larger HO basis ($N_0=16$) in the illustrative case of the 
pairing window parameters $\Delta\epsilon=6$ MeV and $\mu=0.2$ MeV. The 
optimal value of $G$ and the associated root-mean-square error turns 
out to be very close to that obtained with $N_0=10$. This justifies 
the choice of $N_0=10$.

It is worth adding that the authors of Ref.~\cite{Bonche85} performed 
similar calculations in the same mass region (including for $^{76}$Sr and 
$^{80}$Zr) with the SIII Skyrme interaction and the following set of 
parameters for the seniority force: $\Delta\epsilon=5$~MeV, 
$\mu=0.5$~MeV, $G_0^{(n)}=13.5$~MeV and $G_0^{(p)}=16.5$~MeV. 
The pairing strengths were determined from the experimental 
quasiparticle energies extracted in the same way as we did 
from the experimental binding energies. 

\begin{table*}[t] 
\begin{center} 
\caption{Ground-state properties obtained with the SIII and SLy4 Skyrme 
interactions and the pairing strengths obtained through the 3-point and 
the 5-point formulae adjustment procedure. All the quantities in the columns 
at the right of $Q_{40}$ are expressed in MeV. 
\label{res_GS}} 
\vspace*{0.25cm} 
\begin{tabular}{cccccccccccccccc} 
\hline\hline 
& Force & Formula & Nucleus & $\beta_2$ & $r_m$(fm) & $Q_{20}$(b) &   
$Q_{40}$(b$^2$) & $E/A$ & $\Delta_n$ & $\Delta_p$ & $\lambda_n$ &   
$\lambda_p$ & $\mathcal{E}_n$ & $\mathcal{E}_p$ & \\ 
\hline 
& SIII & 3-point & $^{64}$Ge & 0.200 & 3.917 & 2.651 & 0.0097 & 8.4216   
& 1.375 & 1.138 & -12.642 & -2.713 & 1.376 & 1.140 & \\ 
& SIII & 3-point & $^{68}$Se & -0.267 & 4.015 & -3.324 & 0.0535 &   
8.3698 & 1.173 & 0.891 & -13.040 & -2.760 & 1.502 & 1.222 & \\ 
& SIII & 3-point & $^{72}$Kr & -0.340 & 4.117 & -4.550 & 0.1038 &   
8.3229 & 0.966 & 0.579 & -13.057 & -2.479 & 1.299 & 1.081 & \\ 
& SIII & 3-point & $^{76}$Sr & 0.390 & 4.238 & 7.498 & 0.2353 & 8.3024   
& 0.425 & 0.000 & -13.653 & -2.584 & 1.132 & 0.971 & \\ 
& SIII & 3-point & $^{80}$Zr & 0.401 & 4.318 & 8.441 & 0.1879 & 8.2592   
& 0.967 & 0.623 & -13.131 & -1.586 & 1.185 & 0.965 & \\ 
\cline{3-15}
& SIII & 5-point & $^{64}$Ge & 0.191 & 3.918 & 2.515 & 0.0141 & 8.4485   
& 2.035 & 1.690 & -12.775 & -2.830 & 2.036 & 1.690 & \\ 
& SIII & 5-point & $^{68}$Se & -0.244 & 4.012 & -3.062 & 0.0460 &   
8.3967 & 2.074 & 1.709 & -13.119 & -2.807 & 2.238 & 1.876 & \\ 
& SIII & 5-point & $^{72}$Kr & -0.259 & 4.089 & -3.546 & 0.0544 &   
8.3472 & 2.064 & 1.710 & -13.228 & -2.477 & 2.065 & 1.712 & \\ 
& SIII & 5-point & $^{76}$Sr & 0.381 & 4.234 & 7.301 & 0.2133 & 8.3087   
& 1.380 & 0.947 & -13.537 & -2.515 & 1.683 & 1.330 & \\ 
& SIII & 5-point & $^{80}$Zr & 0.384 & 4.308 & 8.031 & 0.1995 & 8.2741   
& 1.766 & 1.387 & -13.256 & -1.685 & 1.828 & 1.495 & \\ 
\cline{2-15}
& SLy4 & 3-point & $^{64}$Ge & 0.166 & 3.887 & 2.141 & 0.0010 & 8.4719   
& 1.162 & 0.978 & -12.687 & -2.831 & 1.162 & 0.979 & \\ 
& SLy4 & 3-point & $^{68}$Se & -0.256 & 3.996 & -3.172 & 0.0344 &   
8.4347 & 0.672 & 0.133 & -12.984 & -2.768 & 1.551 & 1.322 & \\ 
& SLy4 & 3-point & $^{72}$Kr & -0.169 & 4.047 & -2.359 & -0.0038 &   
8.3741 & 1.275 & 1.058  & -13.237 & -2.541 & 1.295 & 1.075 & \\ 
& SLy4 & 3-point & $^{76}$Sr & 0.392 & 4.218 & 7.470 & 0.2380 & 8.3402   
& 0.000 & 0.000 & -13.627 & -2.597 & 1.226 & 1.149 & \\ 
& SLy4 & 3-point & $^{80}$Zr & 0.000 & 4.158 & 0.000 & 0.0000 & 8.3377   
& 0.461 & 0.000 & -13.911 & -2.352 & 1.458 & 1.238 & \\ 
\cline{3-15}
& SLy4 & 5-point & $^{64}$Ge & 0.150 & 3.887 & 1.922 & 0.0035 & 8.4908   
& 1.687 & 1.426 & -12.764 & -2.907 & 1.693 & 1.430 & \\ 
& SLy4 & 5-point & $^{68}$Se & -0.228 & 3.992 & -2.844 & 0.0288 &   
8.4489 & 1.736 & 1.376 & -13.064 & -2.823 & 2.153 & 1.837 & \\ 
& SLy4 & 5-point & $^{72}$Kr & -0.164 & 4.048 & -2.297 & -0.0010 &   
8.3958 & 1.839 & 1.533 & -13.296 & -2.608 & 1.852 & 1.543 & \\ 
& SLy4 & 5-point & $^{76}$Sr & 0.000 & 4.097 & 0.001 & 0.0000 & 8.3614   
& 1.805 & 1.530 & -13.952 & -2.863 & 1.867 & 1.572 & \\ 
& SLy4 & 5-point & $^{80}$Zr & 0.000 & 4.158 & 0.000 & 0.000 & 8.3450   
& 1.365 & 1.065 & -13.931 & -2.383 & 1.971 & 1.655 & \\ 
\hline\hline 
\end{tabular} 
\end{center} 
\end{table*}
With the optimal values of the pairing strength of Table~\ref{Gopt}
we calculate several GS properties related to the nuclear size
(through the root-mean-square mass radius $r_m$) and deformation 
(through $\beta_2$ and the mass quadrupole $Q_{20}$ and hexadecapole 
$Q_{40}$ moments), the binding energy per nucleon ($E/A$) as well 
as pairing quantities (the BCS pairing gaps $\Delta_n$ and $\Delta_p$, 
the chemical potentials $\lambda_n$ and $\lambda_p$, and the 
minimal quasiparticle energies $\mathcal{E}_n$ and $\mathcal{E}_p$). 
The definitions of $\beta_2$, $r_m$, $Q_{20}$ and $Q_{40}$ can be
found in the Appendix. The obtained results are reported in 
Table~\ref{res_GS}. The most striking difference between the two
Skyrme interactions is that they yield very different GS deformations 
for $^{80}$Zr (strongly prolate with SIII, spherical with
SLy4). Moreover the GS deformation of $^{76}$Sr drastically depends on
the pairing strength when using the SLy4 interaction.

Since the GS deformations calculated with the SIII interaction are in
agreement with the experimental data, especially for 
$^{76}$Sr~\cite{Nacher04} and $^{80}$Zr~\cite{Lister01}, we perform
HTDA calculations only with this interaction.

%
%

\section{Results of HTDA calculations}

\label{res_HTDA}

\subsection{Convergence of the HTDA solutions}

In the HTDA calculations two types of truncation schemes need to be 
defined. The first one is concerned with the maximal order in the 
many-particle many-hole basis, the second one with the single-particle
states from which this basis is built. In the latter case, we are
facing thus a situation met in customary BCS calculations. Typically,
we limit our single-particle subspace to the configuration-space 
window defined as
\begin{equation}
|\epsilon_k-\epsilon_F|\leqslant \Delta\epsilon  \,,     
\end{equation}
where $\epsilon_F$ is the Fermi energy and $\Delta\epsilon$ is 
a cut-off energy. The actual value of $\Delta\epsilon$ is chosen 
such that single-particle states left out would not contribute 
significantly except by a renormalization of quantities measuring the 
intensity of pairing correlations (such as the correlation energy defined 
below). As already used in HTDA calculations for the $^{64}$Ge 
nucleus~\cite{Sieja07}, we retain here $\Delta\epsilon=12$ MeV for 
both charge states. As in the BCS treatment, the two-body matrix 
elements are multiplied by the smooth cut-off factor of 
Eq.~(\ref{smooth_cut_off}) with $\Delta\epsilon=12$ MeV and $\mu=0.2$.

A detailed study of the particle-hole expansion of the HTDA
ground state was performed in Ref.~\cite{Pillet05} in the picket-fence
model with 8 or 16 levels filled with 8 or 16 particles,
respectively. The authors obtained the convergence of GS solutions
toward the exact solution of the Richardson model for a
6p-6h space in the standard nuclear pairing regime.  
Nevertheless, it may be expected that the realistic character of the
HF vacuum calculated with a realistic effective interaction
ensures a faster convergence of the particle-hole expansion. This is
what the studies of Refs.~\cite{Pillet02,Sieja07} using
a 4p-4h space tend to indicate.

Within the above truncation scheme for the configuration-space 
window obtained in the HFBCS calculations with the SIII force and
$G=20.6$~MeV, we study the convergence of the HTDA solutions as a function
of the many-body basis size. The three quantities under consideration are 
i) the correlation energy defined as the difference between the
expectation values of the Hamilton operator evaluated in the
correlated and uncorrelated states
\begin{equation}
E_{\rm corr}=\langle\Psi|\hat H\ket{\Psi}-\langle\Phi_0|\hat
H|\Phi_0\rangle\:,
\label{ecorr-def}
\end{equation}
ii) the occupation probabilities $v_i^2$ which are defined in the HTDA
approach as the diagonal matrix element of the one-body density 
$\hat{\rho}$ in the single-particle basis
\begin{equation}
v_i^2=\rho_{ii}=\elmx{\Psi}{\ac{i}a_i}{\Psi}\:,
\label{vi}
\end{equation}
where $\ac{i}$ and $a_i$ are the creation and annihilation operators
associated with the single-particle state $\ket{i}$, respectively, and 
iii) the mass quadrupole moment (see the Appendix for definitions). 
Since the one-body density is not diagonal in the HF basis 
a transformation to the canonical basis is done to obtain the 
$v_i^2$ values.

\begin{figure}
\centerline{\includegraphics[scale=0.6]{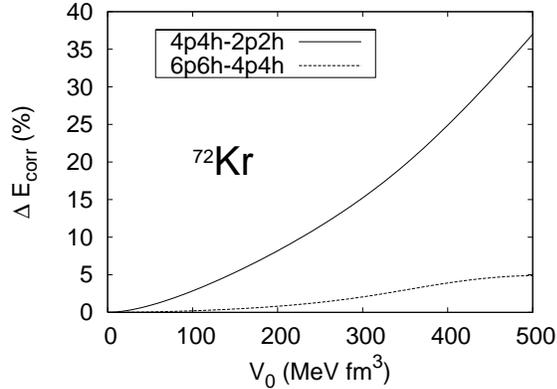}}
\caption{Convergence of the total correlation energy
for $^{72}$Kr as a function of the pairing strength $V_0$.
The percentage difference between correlation energies
obtained in many-body spaces that differ by 2p-2h are shown.\label{econv}}
\end{figure}

The percentage difference of $E_{\rm corr}$ between solutions obtained
in 2p-2h, 4p-4h and 6p-6h spaces of pair excitations are shown
for $^{72}$Kr as a function of the strength $V_0$ of the residual $\delta$
interaction in Fig.~\ref{econv}.
The difference between 2p-2h and 4p-4h solutions is found to be 
very large and increases nearly linearly with the pairing strength $V_0$. 
The discrepancy between 4p-4h and 6p-6h solutions reaches 5\% 
in the strong pairing region. A similar behavior is found for the
other nuclei.

\begin{figure*}[t]
\centerline{\includegraphics[width=\textwidth]{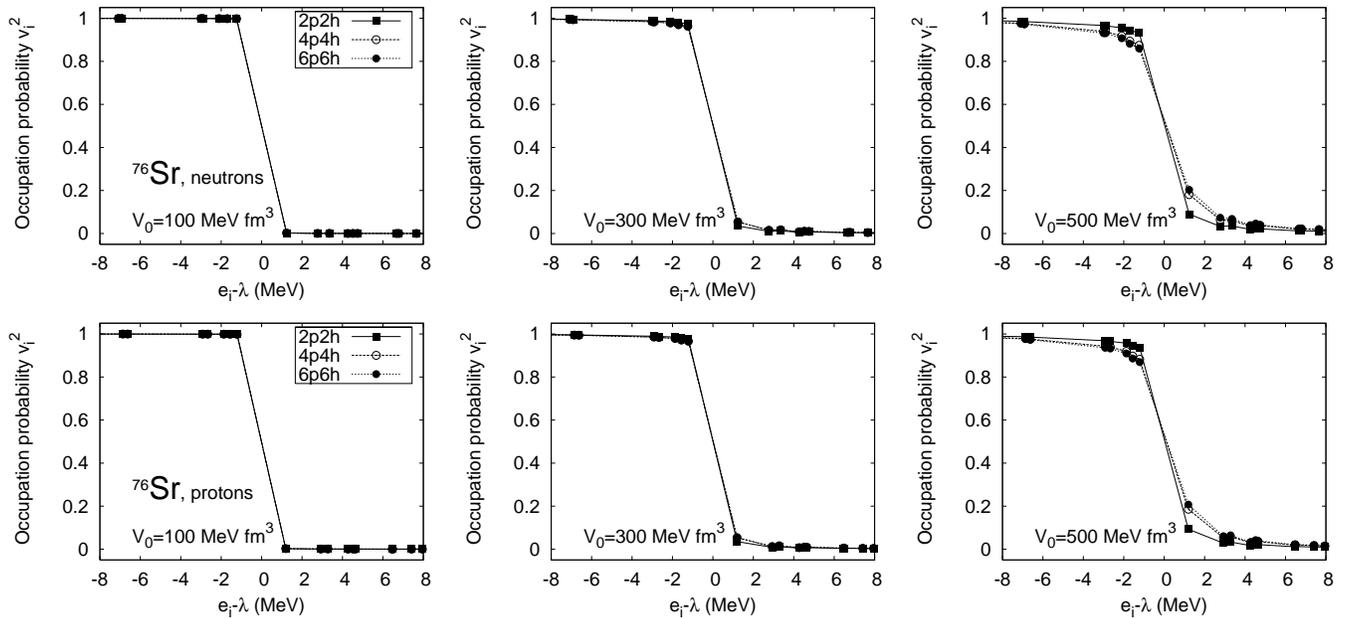}}
\caption{Convergence of the occupation probability
$v_i^2$ with enlarging of the particle-hole excitations space for neutrons
and protons in $^{76}$Sr. The cases of three different intensities
$V_0$ of pairing interaction are shown.\label{v2conv}}
\end{figure*}

The occupation probabilities of single-particle levels obtained in
the different spaces and with different $V_0$-values are shown in
Fig.~\ref{v2conv} for $^{76}$Sr. 
There are no conspicuous differences
between the various solutions except in the stronger pairing case, where
the difference between the $v_i^2$ values in 2p-2h and larger spaces 
becomes substantial. Therefore, the calculated quantities like 
quadrupole moments and radii converge quickly with the maximal order 
of particle-hole excitations. 

In Tables \ref{conv_1} and \ref{conv_2}, the values of correlation
energies and quadrupole moments obtained in 2p-2h, 4p-4h and 6p-6h
spaces are given for all nuclei and for two realistic values of the
strength of the residual interaction, namely $V_0=400$~$\rm MeV.fm^3$ 
and $V_0=320$~$\rm MeV.fm^3$ (see the next subsection for details).
\begin{table}[h]
\caption{Correlation energy and mass quadrupole moments
obtained in the calculations using 2p-2h, 4p-4h and 6p-6h spaces
with the $\delta$-force strength $V_0=400$~$\rm
MeV.fm^3$.\label{conv_1}}
\vspace*{0.25cm}
\begin{tabular}{cc|ccc|cccc}
\hline\hline
& \multirow{2}{1.3cm}{\centerline{Nucleus}} &
\multicolumn{3}{c|}{$E_{\rm corr}$(MeV)} &
\multicolumn{3}{c}{$Q_{20}$(b)} &  \\ 
&           & 2p-2h & 4p-4h & 6p-6h & 2p-2h & 4p-4h & 6p-6h&\\ 
\hline
& $^{64}$Ge &$-4.763$&$-6.383$&$-6.556$&$2.670$&$2.645$&$2.641$&\\
& $^{68}$Se &$-4.127$&$-5.060$&$-5.205$&$-3.301$&$-3.306$&$-3.306$&\\
& $^{72}$Kr &$-4.306$&$-5.348$&$-5.594$&$-4.575$&$-4.528$&$-4.519$&\\
& $^{76}$Sr &$-4.010$&$-4.996$&$-5.042$&$7.493$&$7.482$&$7.480$&\\
& $^{80}$Zr &$-4.077$&$-5.046$&$-5.228$&$8.453$&$8.449$&$8.448$&\\
\hline\hline 
\end{tabular}
\end{table}
\begin{table}
\caption{Same as in Table \ref{conv_1} with the strength 
$V_0=320$~$\rm MeV.fm^3$ of the $\delta$-force.\label{conv_2}} 
\vspace*{0.25cm}
\begin{tabular}{cc|ccc|cccc}
\hline\hline
& \multirow{2}{1.3cm}{\centerline{Nucleus}} &
\multicolumn{3}{c|}{$E_{\rm corr}$(MeV)} &
\multicolumn{3}{c}{$Q_{20}$(b)} &  \\  
 &          & 2p-2h & 4p-4h & 6p-6h & 2p-2h & 4p-4h & 6p-6h&\\ 
\hline
 &$^{64}$Ge &$-2.992$&$-3.796$&$-3.870$&$2.676 $&$2.656 $&$2.655$&\\
 &$^{68}$Se &$-2.404$&$-2.798$&$-2.842$&$-3.301$&$-3.304$&$-3.304$&\\
 &$^{72}$Kr &$-2.558$&$-2.936$&$-3.026$&$-4.595$&$-4.574$&$-4.571$&\\
 &$^{76}$Sr &$-2.493$&$-2.696$&$-2.758$&$7.498 $&$7.493 $&$7.493$&\\
 &$^{80}$Zr &$-2.421$&$-2.768$&$-2.825$&$8.456 $&$8.456 $&$8.456$&\\
\hline\hline 
\end{tabular}
\end{table}
The largest discrepancy between 4p-4h and 6p-6h results
occurs for $^{72}$Kr with $V_0=400$~$\rm MeV.fm^3$ where 
the difference for the correlation energy reaches $3.6\%$. With the
pairing strength $V_0=320$~$\rm MeV.fm^3$ the discrepancies for 
$E_{\rm corr}$ amount on the average to $1\%$. For all five nuclei the 
differences between the values of the quadrupole moments obtained 
in 4p-4h and 6p-6h calculations are negligible.
 
These results suggest that it is reasonable to truncate the
particle-hole expansion at order 4. As a matter of fact, this allows
for a satisfactory accuracy of the calculations at a rather low cost
in terms of computation time. A typical number of many-body
configurations to be handled for each charge state is about $5000$ 
while it may rise typically up to $50000$ or higher when including
6p-6h excitations. It is worth adding here that in these calculations no
additional cut-off on the maximal particle-hole energy of the
scattered pairs is applied. The addition of such a cut-off for
the particle-hole excitation energy would yield an even faster
convergence without losing any substantial accuracy~\cite{Pillet05} 
and may be one way to handle calculations of higher particle-hole
excitation order.

\subsection{Adjustment of the strength of the residual $\delta$-interaction}

The fit of the strength of the $\delta$ interaction in the HTDA approach 
is performed in the space of up to two-pair excitations.
Two adjustment schemes are considered: one where the same strength 
is retained for neutrons ($V_{0n}$) and protons ($V_{0p}$), 
the other one where $V_{0p}$ is reduced by $10\%$ with respect to 
$V_{0n}$ to account for the Coulomb anti-pairing effect as in the HFBCS 
calculations. 

In Table~\ref{fit_htda} we present the optimal $V_{0p}$-values, 
noted $V_0^{\rm opt}$, and the associated gap root-mean-square 
deviations $\sigma_{\Delta}$ in each
adjustment scheme and using the 3-point and 5-point formulae. 
Since the reduction of $V_0^{\rm opt}$ slightly improves the quality
of the fit, we choose this adjustment scheme in further HTDA
calculations.
\begin{table}[h]
\caption{Optimal values $V_0^{\rm opt}$ (in $\rm MeV.fm^3$) and
corresponding root-mean-square errors $\sigma_{\Delta}$ (in MeV) on
pairing gaps obtained with the SIII Skyrme interaction using the 3-point
and the 5-point formulae.
\label{fit_htda}}
\begin{center}
\begin{tabular}{ccccccccc}
\hline\hline
& \multirow{2}{1.3cm}{\centerline{Formula}} & & 
\multicolumn{2}{c}{$V_{0p}=V_{0n}$} & & \multicolumn{2}{c}{$V_{0p}=0.9V_{0n}$} 
& \\
& & & $V_{0}^{\rm opt}$ & $\sigma_{\Delta}$ & & $V_{0}^{\rm opt}$ & 
$\sigma_{\Delta}$ & \\
\hline
& 3-point &&320 &0.278&& 340 & 0.264&\\
& 5-point &&400 &0.285&& 420 & 0.249&\\
\hline\hline
\end{tabular}
\end{center}
\end{table}

\subsection{Ground-state properties obtained within the HTDA approach}

As mentioned in Sect.~\ref{res_HFBCS}, the HTDA calculations use as 
a starting point the HFBCS solutions obtained with the SIII
effective interaction and two sets of values of the pairing-strength
adjusted to experimental data through the 3-point and 5-point
formulae. The results for the GS properties are presented in
Tables~\ref{htda_GS_3} and \ref{htda_GS_5} when using the 3-point and
5-point formulae, respectively. 
\begin{table}[h]
\begin{center}
\caption{Ground-state properties within the HTDA framework with the
SIII force and the pairing strength obtained through the 3-point
formula adjustment procedure. The mass root-mean-square radius
$r_m$, quadrupole moment and hexadecapole moment are given in fm,
barns (b) and $\rm b^2$, respectively, whereas the binding energy per
nucleon $E/A$ and the pairing gaps are expressed in
MeV. \label{htda_GS_3}}
\vspace*{0.25cm}
\begin{tabular}{ccccccccccccc}
\hline\hline
& Nucleus & $\beta_2$ & $r_m$ & $Q_{20}$ & $Q_{40}$ & $E/A$ &
$\Delta_n$ & $\Delta_p$ & \\ 
\hline
\hline
& $^{64}$Ge & 0.201&3.917 & 2.661 &0.0076&8.4340& 1.801 & 1.698&\\
& $^{68}$Se &-0.267&4.014 &-3.315 &0.0530&8.4004& 1.338 & 1.324&\\
& $^{72}$Kr &-0.341&4.120 &-4.575 &0.1059&8.3577& 1.164 & 1.152&\\
& $^{76}$Sr & 0.389&4.240 & 7.494 &0.2344&8.3344& 2.003 & 1.524&\\
& $^{80}$Zr & 0.400&4.320 & 8.444 &0.1830&8.2863& 1.216 & 1.011&\\
\hline\hline
\end{tabular}
\end{center}
\end{table}
\begin{table}[h]
\begin{center}
\caption{Same as Table~\protect\ref{htda_GS_3} using the 5-point formula 
adjustment procedure. \label{htda_GS_5}}
\vspace*{0.25cm}
\begin{tabular}{cccccccccccc}
\hline\hline
& Nucleus & $\beta_2$ & $r_m$ & $Q_{20}$ & $Q_{40}$ & $E/A$ &
$\Delta_n$ & $\Delta_p$ & \\ 
\hline
\hline
& $^{64}$Ge & 0.200&3.919 & 2.651 &0.0089&8.4723& 2.113 & 2.011&\\
& $^{68}$Se &-0.266&4.017 &-3.318 &0.0536&8.4326& 2.023 & 2.006&\\
& $^{72}$Kr &-0.337&4.121 &-4.532 &0.1027&8.3898& 1.718 & 1.714&\\
& $^{76}$Sr & 0.382&4.242 & 7.483 &0.2330&8.3620& 2.003 & 1.524&\\
& $^{80}$Zr & 0.400&4.321 & 8.436 &0.1879&8.3138& 1.513 & 1.323&\\  
\hline\hline
\end{tabular}
\end{center}
\end{table}

Since we are considering here only the equilibrium
deformations as determined by the HFBCS calculations, HTDA 
calculations do not affect much the bulk observables, with the
exception of the binding energy and related quantities such as 
the pairing gaps. The former is indeed much lower (by 2 to 4~MeV) 
for the HTDA solutions than for the HFBCS ones. Regardless of the 
adjustment scheme for the pairing strength, the binding energy 
calculated in the HTDA approach agrees better with the
experimental values (see Table~\ref{Exp_Eqp}) than in the HFBCS approach.

%
%

\section{Pairing properties in BCS and HTDA approaches}

\label{sec-pair}

To evaluate the amount of pairing correlations one might think of
considering the correlation energy, defined as the difference between
the expectation values of the Hamiltonian in the correlated and
uncorrelated (HF vacuum) solutions. However, such a
correlation energy has no realistic character in practice, in both the
HFBCS and HTDA approaches. Indeed, there is no consistency between 
the interaction used to generate the mean-field and the one building
up pairing correlations. Instead, a quantity only related to the
residual interaction would be more significant in this respect. This
is the case of the so-called condensation energy $E_{\rm cond}$. 
In the HFBCS approach, $E_{\rm cond}$ is proportional to the trace of
the product of the abnormal density and the pairing field.
In the HTDA approach, we may define it by
\begin{equation}
E_{\rm cond}=E_{\rm corr}-\sum_i\chi_i^2 E_{\rm p-h}^{i}\,,
\end{equation}
where the $\chi_i^2$ factors are the probability of the configuration $i$ 
whose unperturbed particle-hole energy is $E_{\rm ph}^{i}$.

\begin{table}[h]
\begin{center}
\caption{Neutron condensation energies and diffuseness of the neutron
Fermi surface obtained in the HFBCS and HTDA approaches. The results
with SIII force and two fits of the strength of the pairing
interaction are given in each case. \label{tab-pair-n}}
\vspace*{0.25cm}
\begin{tabular}{cccccccc}
\hline\hline
& \multirow{2}{1.3cm}{\centerline{Formula}} &
\multirow{2}{1.3cm}{\centerline{Nucleus}} & \multicolumn{2}{c}{HFBCS} & 
\multicolumn{2}{c}{HTDA} \\
& & & $E_{\rm cond}$ & $\sum u_iv_i$ & $E_{\rm cond}$ & $\sum u_iv_i$ & \\
\hline
\hline
&3-point&$^{64}$Ge &$-4.80$ & 3.4 & $-5.27$ &3.2&\\
&3-point&$^{68}$Se &$-3.50$ & 3.0 & $-4.55$ &2.8&\\
&3-point&$^{72}$Kr &$-2.50$ & 2.6 & $-4.68$ &2.9&\\
&3-point&$^{76}$Sr &$-0.50$ & 1.2 & $-4.33$ &2.7&\\
&3-point&$^{80}$Zr &$-2.68$ & 2.8 & $-4.61$ &3.0&\\
&5-point&$^{64}$Ge &$-9.04$ & 4.3 & $-8.31$ &3.7&\\
&5-point&$^{68}$Se &$-9.95$ & 4.6 & $-7.85$ &3.6&\\
&5-point&$^{72}$Kr &$-9.78$ & 4.7 & $-8.16$ &3.7&\\
&5-point&$^{76}$Sr &$-4.53$ & 3.3 & $-7.70$ &3.6&\\
&5-point&$^{80}$Zr &$-7.75$ & 4.4 & $-8.20$ &3.9&\\  
\hline\hline
\end{tabular}
\end{center}
\end{table}
\begin{table}[h]
\begin{center}
\caption{ Same as in Table \ref{tab-pair-n} for protons.\label{tab-pair-p}}
\vspace*{0.25cm}
\begin{tabular}{cccccccc}
\hline\hline
& \multirow{2}{1.3cm}{\centerline{Formula}} &
\multirow{2}{1.3cm}{\centerline{Nucleus}} & \multicolumn{2}{c}{HFBCS} & 
\multicolumn{2}{c}{HTDA} \\
& & & $E_{\rm cond}$ & $\sum u_iv_i$ & $E_{\rm cond}$ & $\sum u_iv_i$ & \\
\hline
\hline
&3-point&$^{64}$Ge &$-3.55$ &3.1&$-3.90$&2.9&\\
&3-point&$^{68}$Se &$-2.22$ &2.5&$-3.15$&2.4&\\
&3-point&$^{72}$Kr &$-1.00$ &1.7&$-3.40$&2.6&\\
&3-point&$^{76}$Sr &$-0.00$ &0.0&$-2.94$&2.3&\\
&3-point&$^{80}$Zr &$-1.42$ &2.1&$-3.01$&2.4&\\
&5-point&$^{64}$Ge &$-6.73$ &3.9&$-6.12$&3.4&\\
&5-point&$^{68}$Se &$-7.34$ &4.2&$-5.58$&3.2&\\
&5-point&$^{72}$Kr &$-7.46$ &4.3&$-6.11$&3.4&\\
&5-point&$^{76}$Sr &$-2.36$ &2.5&$-5.33$&3.1&\\
&5-point&$^{80}$Zr &$-5.70$ &4.0&$-5.48$&3.3&\\  
\hline\hline
\end{tabular}
\end{center}
\end{table}
Another variable that may shed light on the amount of pair correlations 
is the trace of the positive-definite operator
$\hat{\rho}^{\frac{1}{2}}(1-\hat{\rho})^{\frac{1}{2}}$. Indeed, this
quantity expresses the non-idempotent character of the density
operator $\hat{\rho}$ and is thus related to correlations. As well
known in the BCS case, it is related to the abnormal density. Using
the occupation factor $v_i$ defined in Eq.~(\ref{vi}), we can write
the trace of the operator
$\hat{\rho}^{\frac{1}{2}}(1-\hat{\rho})^{\frac{1}{2}}$ simply as the 
sum $\sum_i u_iv_i$, with $u_i=\sqrt{1-v_i^2}$.

\begin{table*}[t]
\begin{center}
\caption{Ground-state properties of the five considered nuclei as
functions of the $x$ value (see text for details).
\label{tab-pn}}
\vspace*{0.25cm}
\begin{tabular}{ccccccccccccc}
\hline\hline
& \multirow{2}{1.3cm}{\centerline{Formula}} &
\multirow{2}{1.3cm}{\centerline{Nucleus}} & \multicolumn{3}{c}{$r_m$ (fm)}
& \multicolumn{3}{c}{$Q_{20}$ (b)} & \multicolumn{3}{c}{$Q_{40}$ 
($\rm b^2$)} & \\
& & & $x=0$ & $x=1$ & $x=2$ & $x=0$ & $x=1$ & $x=2$ & $x=0$ & $x=1$ &
$x=2$ & \\
\hline
&3-point&$^{64}$Ge &3.916&3.916&3.917& 2.691& 2.693& 2.697&0.0065&0.0064&0.0062&\\
&3-point&$^{68}$Se &4.012&4.012&4.012&-3.309&-3.308&-3.308&0.0523&0.0522&0.0522&\\
&3-point&$^{72}$Kr &4.119&4.120&4.120&-4.601&-4.602&-4.603&0.1081&0.1082&0.1084&\\
&3-point&$^{76}$Sr &4.239&4.240&4.240& 7.499& 7.501& 7.506&0.2354&0.2356&0.2360&\\
&3-point&$^{80}$Zr &4.320&4.320&4.321& 8.446& 8.446& 8.450&0.1798&0.1795&0.1792&\\
&5-point&$^{64}$Ge &3.917&3.918&3.918& 2.688& 2.690& 2.696&0.0069&0.0067&0.0065&\\
&5-point&$^{68}$Se &4.013&4.013&4.013&-3.308&-3.308&-3.307&0.0525&0.0524&0.0523&\\
&5-point&$^{72}$Kr &4.120&4.120&4.120&-4.590&-4.592&-4.490&0.1073&0.1075&0.1078&\\
&5-point&$^{76}$Sr &4.240&4.240&4.240& 7.495& 7.498& 7.506&0.2351&0.2354&0.2361&\\
&5-point&$^{80}$Zr &4.321&4.321&4.322& 8.442& 8.444& 8.450&0.1811&0.1806&0.1801&\\  
\hline\hline
\end{tabular}
\end{center}
\end{table*}
These two measures of pairing correlations calculated in the HFBCS and 
HTDA approaches are compared separately for neutrons and protons in 
Tables~\ref{tab-pair-n} and \ref{tab-pair-p}, respectively. 
The most striking feature is the rather tiny variations in the HTDA 
case from one nucleus to another. This is due to the resilience of the
HTDA solutions to react on variations of the level density at the
Fermi surface. In contrast, it is overemphasized in the HFBCS
calculations. One example of this is to be found for the protons
distribution in $^{76}$Sr, where in the HFBCS calculations with a
$G$-value fitted to the 3-point pairing indicator, no superfluid
solution is found (see also Table~\ref{res_GS} for the values
of pairing gaps).

%
%

\section{Proton-neutron pairing in the HTDA approach}

\label{pn_pairing}

In spite of the wide recent theoretical interest paid to the
proton-neutron pairing mode, it is still uncertain what is the exact
importance of its $T=0$ component. Moreover, its connection with the
Wigner energy is not completely clarified and there are a lot of
controversies about other signatures. Since the $T=0$ pairing is
neglected in all fits of the effective interactions in use to
calculate the mean-field, this missing contribution may cause some
artificial bias in the outcome. In the calculations of masses that
make use of a macroscopic energy (liquid-drop models), this
problem is circumvented by adding an ad hoc Wigner term. It is however
highly predictable that the inclusion of proton-neutron pairing 
correlations within a HFB approach would lead to novel features of the
mean field, as for instance different deformation properties since the
usual spin-triplet ($T=0$) pairing mode would tend to break the axial
symmetry. However, since we do not carry out self-consistent
HTDA calculations, the mean field itself is not affected 
by the presence of these proton-neutron correlations.

Expectation values of various observables (such as the radii and
quadrupole moments) specifying the correlated wave function have been
calculated from solutions obtained after diagonalization of a full
isoscalar and isovector residual interaction. For the reasons
mentioned in the previous paragraph, these quantities are not expected
to change much in this perturbative treatment. In contrast, the change
in energy brought in by the consideration of a full $T_z=0$ part of
the interaction is expected to be more significant. In the HTDA 
framework as applied here, and as far as relative variations are 
concerned, the correlation energy defined in Eq.~(\ref{ecorr-def}) 
is relevant. 

The calculations reported in this section are performed in the 4p-4h
space of pair excitations containing all new configurations that
result from the coupling of neutron and proton states to produce the
$0^+$ ground state (including aligned proton-neutron pairs). 
The proton-neutron configurations considerably enlarge
the dimension of the Hamiltonian matrices to be computed and
diagonalized (up to $\sim10^5$). As a result, the computing time 
becomes an issue. This is why we do not test here the convergence of the
particle-hole expansion by going up to order 6 as we have done in the
previous section. A more detailed study of the proton-neutron pair
correlations in the HTDA approach will be given in a forthcoming
publication. In addition, no evidence from previous theoretical
approaches has been found for the $T=0$ collectivity. On the contrary,
a rather strong quenching of the isoscalar pairing has been observed
in the particular case of $pf$-shell nuclei. It is thus likely that we
only have to deal with a somewhat weak neutron-proton pairing and
consequently the 4p-4h limitation should not constitute a stringent
constraint.
    
Based on isospin invariance arguments we choose the strength
$V_{0pn}^{T=1}=1/2(V_{0p}^{T=1}+V_{0n}^{T=1} )$ for the part of the 
residual interaction acting on neutron-proton two-body states. Since
the actual $T=0$ pairing strength is unknown, in other words since
one does not know what should be the data pertaining 
to the determination of a phenomenological such interaction, we adopt
an exploratory approach by varying the ratio $x=V_{0}^{T=0}/V_{0}^{T=1}$ 
of the residual interaction in the two isospin channels from 0.5 to 
2.0 by steps of 0.5.

In Table~\ref{tab-pn} the resulting GS deformations and radii are
indicated only for three values of $x$, namely 0, 1 and 2. 
As expected, these quantities do not vary significantly with $x$.

\begin{figure}
\includegraphics[scale=0.6]{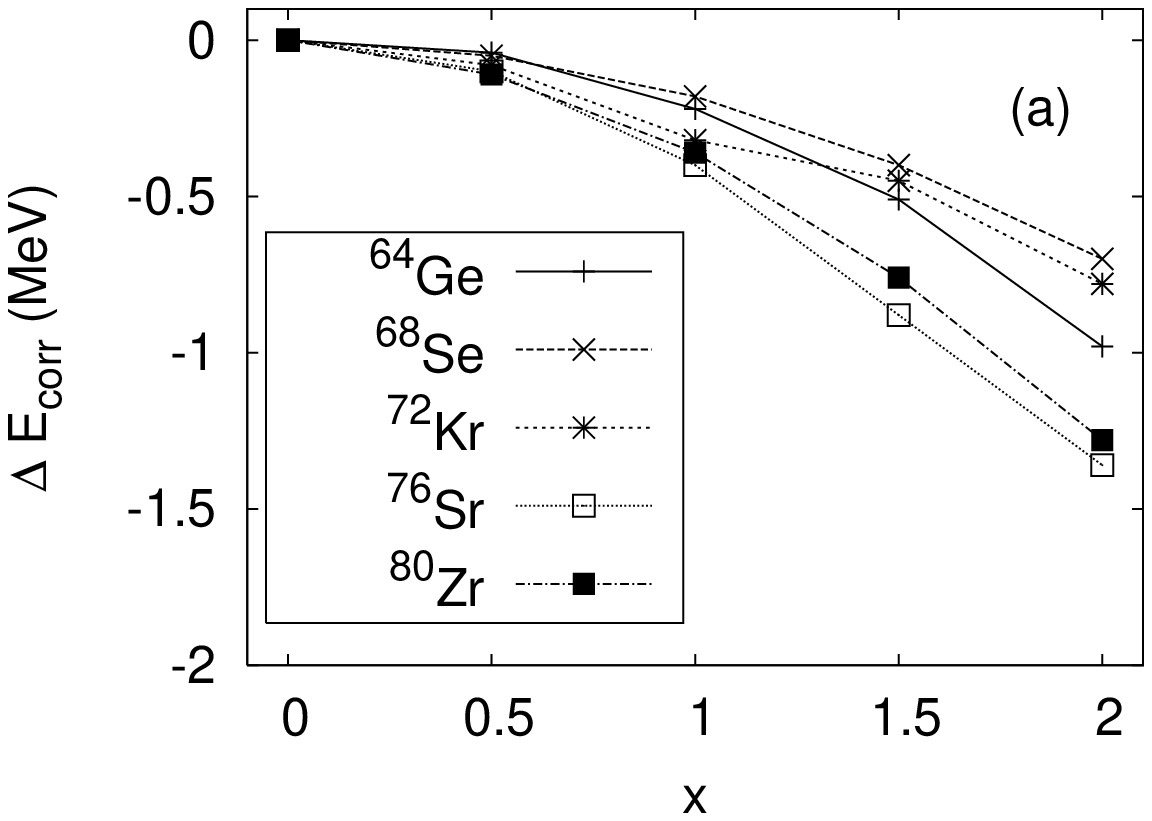}
\includegraphics[scale=0.6]{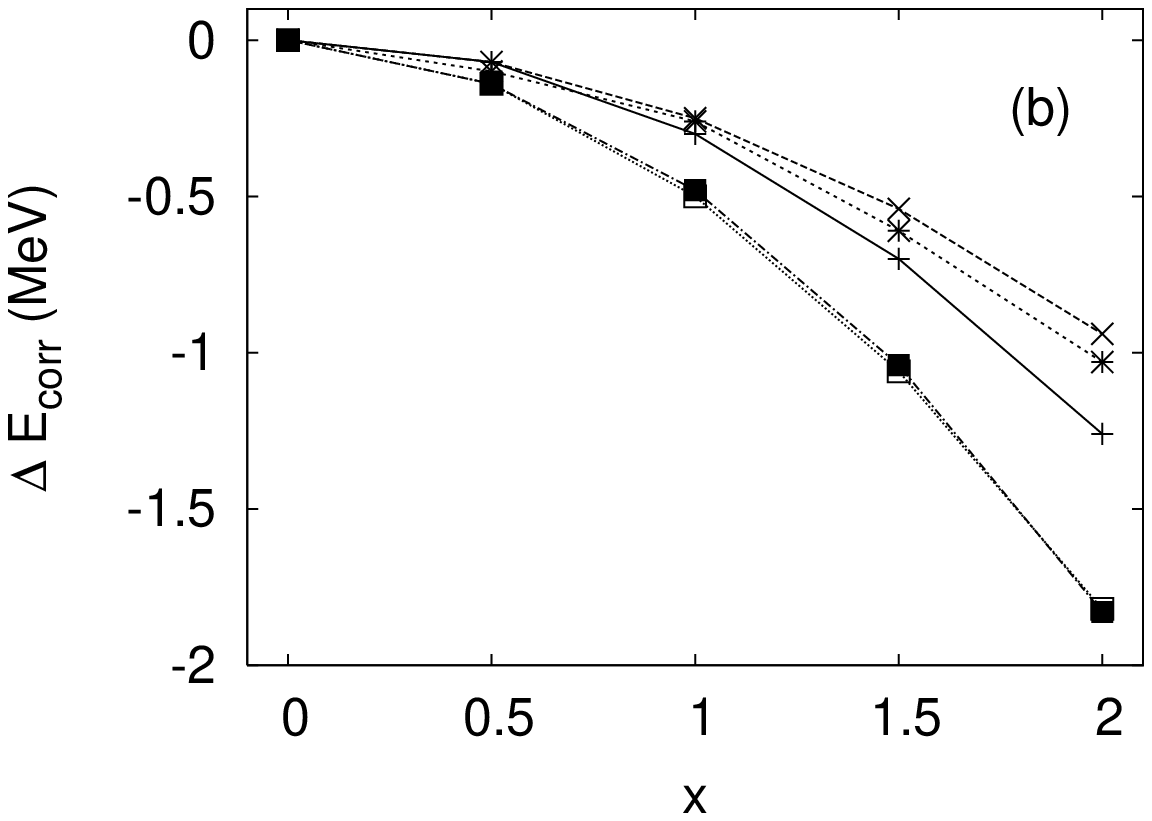}
\caption{Correlation energy (in MeV) normalized to the solution without
$T=0$ pair correlations ($x=0$) as a function of $x$ for all
considered nuclei. The case (a), respectively (b), corresponds to
calculations with the pairing strength adjusted through the 3-point
formula, respectively 5-point formula, in the $T=1$
channel.\label{fig-corr-pn}}
\end{figure}

The relative correlation energy with respect to the $T=0$ pairing
mode, i.e., the difference between the values of 
$E_{\rm corr}$ calculated with a given $x$ value and with $x=0$, is
plotted in Fig.~\ref{fig-corr-pn} as a function of $x$.
The correlation energy induced by the $T=0$ mode is rather important
for large $x$ values, so it is highly desirable to get some
guidance on the actual value of $x$. One could argue that the
shell-model estimates of Ref.~\cite{Dufour96} provide a value of about
$1.5$. However, the shell-model pairing definition with only $(J,T)=(0,1)$
and $(J,T)=(1,0)$ couplings is not consistent with the HTDA one where the
$\delta$-force in use includes all multipolarities. 
Therefore, a systematic comparative
study of pair correlations in large shell-model calculations and in
the HTDA approach would be very helpful to better assess a
realistic value of $x$.

Finally it is important to add that the energy shift due to the $T=0$
pair correlations depends on the pairing intensity in the $T=1$ channel. 
Indeed these pair correlations are larger when the $T=1$ pairing interaction 
is stronger. This suggests that, in a full treatment of these correlations, 
the adjustment of the pairing parameters in the $T=1$ channel should be 
done with proton-neutron correlations.

%
%

\section{Conclusions and perspectives}

\label{conclusion}

The purpose of this paper is to provide a firm basis to study 
the correlations present in deformed even-even $N=Z$ nuclei in the 
$A\sim{70}$ region. We mean, first of all, pairing
correlations in the sense of Cooper pair excitation of nucleons belonging 
to similar orbits. For these particular nuclei, 
proton-neutron correlations must clearly be included in the scheme. 
Moreover, RPA correlations should be further added. In view of the 
difficulties to be overcome in achieving this program, we have undertaken 
here initial steps by including only proton-neutron pairing correlations 
in an exploratory way.

Since the various traditional pairing schemes often lead to a weak coupling
regime, the Bogoliubov vacuum ansatz for the correlated ground state is 
rather inadequate. This is especially true for the not-so-heavy nuclei 
considered here. 
%
%
In contrast the HTDA framework 
represents a consistent and more physical way of handling various kinds 
of correlations, including RPA correlations, on the same footing. 
In the present work, we have chosen to use it from as good as possible 
HFBCS solutions with the usual pairing treatment which involves the 
$T=1$ channel only.

We have had, therefore, to carefully study the pairing channel 
which includes in particular the choice of a $\delta$ interaction 
strength consistent with the data on atomic masses in the studied 
region. As a result, we have calculated some deformation properties 
which are naturally of paramount importance to grasp the single-particle 
spectroscopic properties. Upon determining the HTDA correlated ground 
state, we have shown that the diffuseness of the Fermi surface is 
somewhat similar to what HFBCS yields but less sensitive to the 
fluctuations of the level density as a function of the mass number.

In the last step, where the full $T_z = 0$ residual interaction has 
been considered, we have established the feasibility of such 
calculations. Then, we have assessed 
in a quantitative way how much the relative amount of $T = 0$ and $T = 1$ 
components of the residual interaction influences the correlations 
properties. Even though in this study the $T =1$ interaction strength 
is the one adjusted in the absence of a $T_z = 0$ component, it clearly 
appears from our results that a determination of the above ratio of 
isospin components is badly needed.

The directions of improvement are therefore easy to perceive. 
First they imply a better understanding of the residual interaction. 
Then, RPA correlations should be included in a consistent way, which 
is currently in progress.

\section*{ACKNOWLEDGEMENTS}

This work has benefited from the support by the U.S. Department of Energy 
under contract W-7405-ENG-36, from a funding obtained within the 
Polish-French (IN2P3) laboratories agreement and from a fellowship of 
the French Embassy at Warsaw to support the co-sponsored thesis work 
of one of us (K.S.). These organizations are gratefully acknowledged.

\section*{\uppercase{Appendix: Specific quantities used to assess 
the deformation properties of our solutions}}

In the HFBCS case, the expectation value of a local one-body operator
may be expressed as a space integral involving the local (diagonal in
$\vec{r}$) one-body reduced density matrix. For the mass or isoscalar
(neutron plus proton) distribution, one defines the root-mean-square
radius $r_m$ and the quadrupole $Q_{20}$ and hexadecapole $Q_{40}$
moments as 
\begin{gather}
r_m=\sqrt{\frac{\int d^3\mathbf{r}\,\rho(\mathbf{r})\, \mathbf{r}^2}{A}}\,,\\
Q_{20}=\int d^3\mathbf{r}\,\rho(\mathbf{r})\,(2z^2-x^2-y^2)\,,\\
Q_{40} =\int d^3\mathbf{r}\,\rho(\mathbf{r})\,r^4\,Y_4^0(\theta)\,,
\end{gather}
where $Y_{\ell}^{0}$ denotes the spherical harmonic of order $\ell$
and magnetic quantum number 0.

We then consider the equivalent spheroid which has the same 
root-mean-square radius and quadrupole moment as the actual nucleus. 
Denoting the semi-axes along the symmetry axis and perpendicular 
to it by $a$ and $c$, respectively, we have
\begin{gather}
A\,r_m^{\:2}=\frac{1}{5}\,(2a^2+c^2)\,,\\
Q_{20}=\frac{2}{5}\,A\,(c^2-a^2)\,.
\end{gather}
The $\beta_2$ parameter is then calculated for this equivalent 
spheroid by expanding the nuclear radius in polar coordinates 
according to the $\beta_l$-parameterization~\cite{Moller95}
\begin{align}
R(\theta)&=\frac{a}{\sqrt{1-\alpha\,\cos^2 \theta}}\\
&=R_0\,\left(1+\sum_{l=1}^{\infty}\beta_l\,Y_l^0(\theta)\right)\:,
\end{align}
with
\eq{
\alpha=1-\frac{a^2}{c^2}\:.
}
This allows us to calculate analytically the expression of $\beta_2$ for 
the equivalent spheroid as a function of $\alpha$ as
\eq{
\beta_2=\begin{cases}
\sqrt{5\pi}\left[\frac{3}{2\alpha}\left(1-\frac{\sqrt{\alpha(1-\alpha)}}{{\rm Arcsin}\, \sqrt{\alpha}}\right)-1\right] & \alpha\in]0;1[ \\
0 & \alpha=0 \\
\sqrt{5\pi}\left[\frac{3}{2\alpha}\left(1-\frac{\sqrt{-\alpha(1-\alpha)}}{{\rm ln}\, (\sqrt{-\alpha}+\sqrt{1-\alpha})}\right)-1\right] & \alpha<0
\end{cases}\:.
}

In the HTDA case, the above quantities have to be evaluated in the 
correlated state $\ket{\Psi}$ and one cannot use anymore the usual 
generalized Wick theorem. Indeed, one has to evaluate matrix 
elements between two Slater determinants, generally different. 
One might keep, however, the Wick theorem framework by using mixed 
densities \`a la L\"owdin \cite{Lowdin55}. Instead, here we reduce these 
many-body matrix elements into matrix elements evaluated between 
single-particle states, which makes the HTDA calculation of the 
above expectation values similar to the HFBCS ones.

\end{document}